\PassOptionsToPackage{linesnumbered,ruled}{algorithm2e}
\documentclass{l4dc2024}[preprint]

\usepackage{cases}
\usepackage{mathtools}
\usepackage{wrapfig}
\usepackage{algpseudocode}
\usepackage{algorithm}

\newtheorem{problem}{Problem}
\newtheorem{assumption}{Assumption}
\SetKwRepeat{Do}{do}{while}%
\title[Direct Data-Driven Discrete-time BBR]{Direct Data-Driven Discrete-time Bilinear Biquadratic Regulator}
\usepackage{times}

 \coltauthor{\Name{Author Name1} \Email{abc@sample.com}\and
  \Name{Author Name2} \Email{xyz@sample.com}\\
  \addr Address}

 \author{
 \Name{Shanelle G. Clarke} \Email{clarke24@purdue.edu}\\
	\addr 701 West Stadium Ave, West Lafayette IN 47907  \\
  \Name{Omanshu Thapliyal} \Email{omanshu.thapliyal@hal.hitachi.com}\\
  \addr{Big Data Analytics \& Solutions Lab, Hitachi America Ltd., Santa Clara CA 95050}\\
  \Name{Inseok Hwang} \Email{ihwang@purdue.edu}\\
  \addr 701 West Stadium Ave, West Lafayette IN 47907}

\begin{document}

\maketitle

\begin{abstract}%
We present a novel direct data-driven algorithm that learns an optimal control policy for the Bilinear Biquadratic Regulator (BBR) for an unknown bilinear system. The BBR is difficult to solve owing to the presence of the nonlinear biquadratic performance index and the bilinear cross-term in the dynamics. To address these difficulties, we apply several transformations on the state decision variables to obtain a nonlinear optimization problem with a linear performance index and affine (in the parameterized control) state-dependent equality. The adroit use of the Hamiltonian and Pontryagin's Minimum Principle allows us to 
derive a pair of first-order necessary conditions that, at each point in time, are easily solvable linear matrix equalities (LMEs) which give the optimal state-dependent control law. We then use the marginal sample autocorrelation of the collected data to obtain a direct data-driven equivalent of these LMEs. We demonstrate the performance of the proposed algorithm via illustrative numerical examples. 
\end{abstract}


\section{Introduction}
Direct data-driven control approaches focus on designing controllers from available input-output data of a dynamical system without explicit identification of the system model. These approaches are important for applications in which identifying a system model may be difficult, computationally expensive, time consuming, and/or the amount or quality of the available data may be poor \citep{tang2022data}. However, most data-driven control methods focus on synthesizing control laws for linear systems where well-established linear system tools and behavioral techniques (see \citep{Willems2005}) have been leveraged to design stabilizing robust and optimal control laws \citep{DePersis2020}, predictive controllers \citep{berberich2020data}, network controllers \citep{allibhoy2020data}, etc. In contrast, there have been fewer works for nonlinear systems.

To contribute to the body of work on synthesizing data-driven control methods for nonlinear systems, we propose to study a special class of nonlinear dynamical systems, stemming from Volterra–Wiener Theory \citep{rugh1981nonlinear}, called\textit{ Bilinear Systems} which are known to act as bridges between linear and nonlinear control theory \citep{bruni1974bilinear}. Of primary interest is the fact that bilinear systems exhibit ``nearly linear'' behavior allowing for the adroit use of well-developed linear system methods and theories \citep{bruni1974bilinear}, while simultaneously acting as close approximators for more general nonlinear systems \citep{Mohler1970}. Many biological (e.g., population control, cell metabolism, etc.), chemical (e.g., enzyme catalytic reactions, chemical reactors, etc.), and physical (e.g., nuclear fission, etc.) processes naturally admit bilinear models \citep{Mohler1970}. Moreover, there have been a recent surge in the design of bilinear control methods for robotic applications~\citep{folkestad2022koopnet} since lifting techniques, such as Koopman Operators~\citep{bruder2021} and Perron-Frobenius~\citep{huang2022convex}, have been used to transform general nonlinear systems into possibly infinite-dimensional bilinear control systems. In fact, control-affine nonlinear systems can be exactly bilinearized, via Koopman lifting operators, under certain conditions \citep{goswami2021bilinearization}.

Model-based methods (in which the model is completely known a priori) for control of bilinear systems are numerous (see \citep{gao2010successive,BITSORIS2008, zhao2016gramian, wang2018free, halperin2021}), while seminal direct data-driven control methods, which include \citep{yuan2022data,bisoffi2020data,guo2021data,markovsky2022data}, are few but increasing. In \citep{bisoffi2020data,guo2021data}, suboptimal linear state feedback controllers are designed using robust control techniques and polyhedral invariant set approaches that treat the system nonlinearities as process disturbances. Meanwhile, the work in \citep{yuan2022data} proposes a two-trial online control experiment, run over some finite time horizon $T$, that obtains a locally optimal control sequence which minimizes a quadratic performance index using $T$-persistently exciting data reparametrized as Hankel matrices with linear time-invariant embeddings.

We focus on the synthesis of a data-driven controller for a new class of optimal control problems for inhomogeneous bilinear systems with a biquadratic performance index, called the  bilinear biquadratic regulator (BBR) as was first presented in \citep{halperin2021}.\begin{wrapfigure}{r}{0.65\textwidth}
\begin{center}
\vspace{-2em}
\includegraphics[width=0.5\columnwidth]{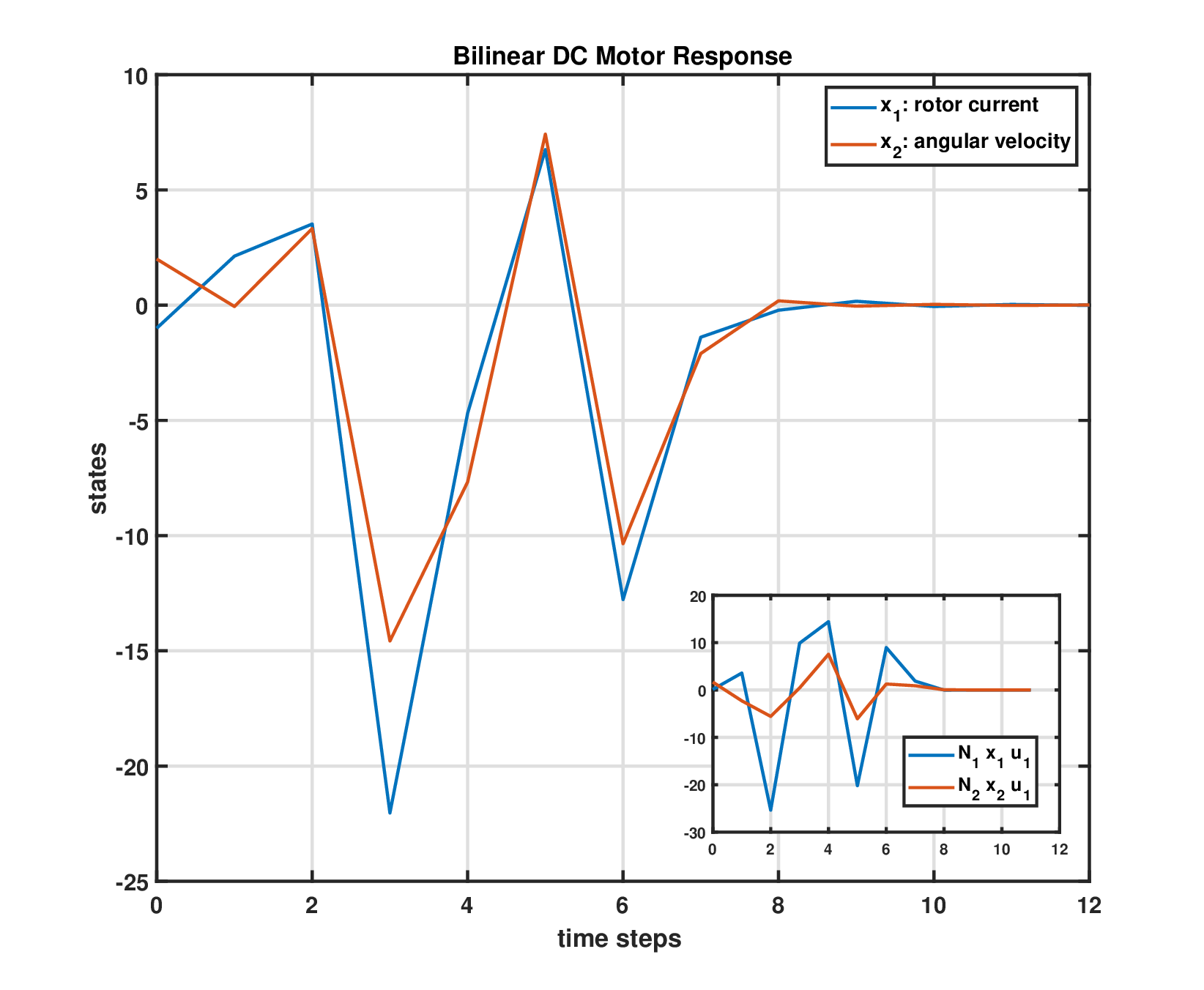}    
\vspace{-1.5em}
\caption{\footnotesize Control of a bilinear DC motor using \citep{wang2018free} with quadratic performance costs: $Q = \mathbb I_2$, $R = 1$.} 
\label{fig:fig1}
\end{center}
\vspace{-2em}
\end{wrapfigure} Consideration of the biquadratic performance index is important for attenuation of the semi-active effects that the control input exerts on the states via the system nonlinearity.  Namely, we want to mitigate any undesirable effects on the closed-loop transient performance produced by the nonlinear state-input coupling (e.g., oscillatory transient behavior in Fig.~\ref{fig:fig1}). While the method in \citep{yuan2022data} generates control laws that drive the state to the origin in finite time $T$, it is unable to comment on the transient system behavior by virtue of its implicit embeddings nor can the convex-concave optimization method therein be trivially modified to accommodate the bilinear nonconvex cross terms introduced by the biquadratic performance index. 

In this paper, we present a novel direct data-driven control algorithm that uses the sample autocorrelation of the state-input data to solve, in a point-to-point manner, a pair of first-order conditions, derived using Pontryagin's Minimum Principle, for an optimal state-dependent nonlinear control law to the discrete-time BBR problem. The main contribution of this paper is the transformation of the BBR with a nonlinear biquadratic performance index and bilinear cross-term in the dynamics, to a nonlinear optimization problem whose solution, point-to-point in time, involves iteratively solving a pair of data-encoded linear matrix equalities. We demonstrate the performance of the proposed algorithm via illustrative numerical examples.

The rest of the paper is organized as follows. Section \ref{section:problem} describes the infinite-horizon bilinear biquadratic regulator for the unknown bilinear system. In Section \ref{section:algorithm}, we derive the direct data-driven optimal control conditions which solves the BBR at instantaneous time increments using Pontryagin's Minimum Principle (PMP). We demonstrate the performance of the proposed algorithm via numerical examples in Section \ref{section:examples_BBR}. Section \ref{section:conclusion} presents our concluding remarks.

\section{Problem Formulation}\label{section:problem}
Consider the discrete-time inhomogeneous bilinear system: 
\begin{align}\label{eqn:bti_plant}
          & x_{t+1} = A x_{t} + B u_{t} + D(x_t \otimes u_{t}), \, t \in \mathbb N_+
\end{align}
where $x_{t} \in \mathbb{R}^n$ and $u_{t} \in \mathbb{R}^m$ are the fully observed system state and input vectors, respectively. $A \in \mathbb{R}^{n \times n} , B \in \mathbb{R}^{n \times m}$, and $D \in \mathbb{R}^{n \times nm} $ are the unknown system matrices. 
\begin{assumption}\label{assumption:span}
The pair $\left(A,\begin{bmatrix} B & D\end{bmatrix}\right)$ is stabilizable.
\end{assumption}

The subspace formed by the set of reachable states of the bilinear system~\eqref{eqn:bti_plant} is not necessarily linear. That is, linear combinations of the reachable states of \eqref{eqn:bti_plant} may not be reachable. Thus, it is difficult to characterize the controllability of~\eqref{eqn:bti_plant} using classical linear algebra tools \citep{rugh1981nonlinear}. Instead, we ask for stabilizability of the pair $(A,\begin{bmatrix} B & D\end{bmatrix})$, a weaker condition than asking for the controllability of~\eqref{eqn:bti_plant}. Note that the work in \citep{yuan2022data} requires controllability of the pair $\left(A,\begin{bmatrix} B & D\end{bmatrix}\right)$. Our method further relaxes this requirement. 

In this paper, we consider the following problem\footnote{\textbf{Relevant Notations:}
 \texttt{blkdiag}$(\cdots)$: a block diagonal matrix; $\lVert \cdot \rVert_2$ is the $\mathcal L_2$ vector norm; $\lVert \cdot \rVert_F$ denotes the Frobenius norm;$\lvert \cdot \rvert$ is the component-wise absolute value of a vector; 
  $\mathbb I_p$ is the $p \times p$ identity matrix; $M\succ 0$, $M\succeq 0$, and $M\preceq 0$ denote a positive definite (PD), positive semi-definite (PSD), and negative semi-definite (NSD) matrix, $M$, respectively. $M_{nn} \in \mathbb{R}^{n \times n}, M_{nm} \in \mathbb{R}^{n \times m}, M_{mn} \in \mathbb{R}^{m \times n}$, and $M_{mm} \in \mathbb{R}^{m \times m}$ are the partitions of the block matrix $M \in \mathbb{R}^{(n+m) \times (n+m)}, M = \begin{bmatrix}
M_{nn} & M_{nm} \\ M_{mn} & M_{mm}\end{bmatrix}$, and  $\otimes$ denotes the kronecker product. $\dagger$ is the Moore-Penrose pseudoinverse of a matrix. $\mathbf 0$ is a matrix (or vector) of zeros with appropriate dimension. $\rho(M)$ denotes the spectral radius of a matrix $M$. \textbf{Tr}$(\bullet)$ denotes the Trace of $\bullet$. Let $\{v_i\}_{i=1}^r \coloneqq \{v_1, \ldots,v_r\}$ denote the set of vectors $v_i$ corresponding to the $i$-th trajectory. $\mathbb N_+$ denotes the set of non-negative natural numbers.
}.
\begin{problem}[Discrete-time Infinite-horizon Bilinear Biquadratic Regulator (BBR)]\label{problem:BBR}
  Consider the discrete-time inhomogeneous bilinear system \eqref{eqn:bti_plant} with the unknown system matrices $(A,B,D)$ under Assumption~\ref{assumption:span}. At each time step $t \in \mathbb N_+$, the state-input transition tuple $(x_t,u_t,x_{t+1})$ from the current state $x_t \in \mathbb R^n$ under some chosen control input $u_t \in \mathbb R^m$ can be fully observed. It is desired to learn a stabilizing state-dependent control gain $K(x_t) \in \mathbb{R}^{m \times n}$, which allows for a nonlinear stabilizing state-feedback control law of the form $u_t = K(x_t) x_t,$ that solves the optimization problem, 
   \begin{align}\label{eqn:problemP0}
    \text{P}_\text{BBR}: 
\begin{cases}
  &  \min\limits_{u_{t}} {J}_0^\infty (x_{t},u_{t}) \coloneqq   \sum\limits_{t=0}^{\infty} \begin{bmatrix}x_{t}  \\ u_{t}  \end{bmatrix}^T \begin{bmatrix} Q &\mathbf 0 \\ \mathbf 0 & R
  \end{bmatrix}\begin{bmatrix}x_{t} \\ u_{t}  \end{bmatrix}  
 + z_t^T  F z_t \\
  &  \text{such that } \eqref{eqn:bti_plant} \text{ holds, } and\lim\limits_{t \to \infty} x_{t} = 0, \,   t \in \mathbb N_+,
\end{cases}
\end{align} 
where $z_{t} \coloneqq  (x_t \otimes u_t)$ is the bilinear cross-term whose effect is to be penalized. We define $Q \succeq 0$, $R \succ 0$, and $F \succeq 0$ as the cost matrices which characterize the biquadratic cost function ${J}_0^\infty (x_{t},u_{t})$.
\end{problem}
We are familiar with the notion of state and actuator control penalties, $Q$ and $R$, respectively from LQR theory~\citep{anderson2007optimal}. In the BBR, the $F$ matrix penalizes the dynamic effect that the control effort exerts on the system states as the bilinear system~\eqref{eqn:bti_plant} evolves. Intuitively, this effect allows that the control law $u_{t}$ actively acts as a damper that dissipates kinetic energy from the structure of~\eqref{eqn:bti_plant}.  Performance indices of the type $J_0^\infty$ naturally arise in vibration attentuation problems \citep{tseng1994semi}, but may also be desirable for many other practical applications. 
We assume the following.
\begin{assumption}\label{assumption:detectability}
$(Q^\frac{1}{2},A)$ is detectable and $(F^\frac{1}{2},D)$ is detectable.
\end{assumption}
Detectability of $(Q^\frac{1}{2},A)$ is standard in LQR theory \citep{anderson2007optimal}. Often taken for granted, this detectability assumption ensures that the state $Q^\frac{1}{2} x_t$ can be determined at each time instance to appropriately reduce $J_0^\infty$. In the same vein, Assumption~\ref{assumption:detectability} allows the state $F^\frac{1}{2} z_t$ to be determined for minimization of $J_0^\infty$. From the work in \citep{halperin2021}, we can solve the model-based version of Problem~\ref{problem:BBR} via Krotov's method \citep{Krotov1993}. Krotov's method optimizes over a sequence of smooth improving functions to obtain a piecewise continuous control sequence such that $J_i^\infty \geq J_{i+1}^\infty $, i.e, $J_0^\infty$ monotonically converges to some optimum. However, the proposed method, based on Krotov's method, in \citep{halperin2021} requires complete model knowledge for solving its improving functions and is computationally complex as the algorithm improves on the smooth improving functions through successive improvements in control over end-to-end simulation trials of horizon $T$.

In what follows, we design a direct data-driven algorithm that learns how to control the unknown bilinear system~\eqref{eqn:bti_plant} by solving Problem~\ref{problem:BBR}, assuming that the system matrices $(A, B, D)$ are unknown, but we have access to the bilinear system~\eqref{eqn:bti_plant} to fully observe the state-input data. 

\section{Main Results and Algorithm}\label{section:algorithm}
Our proposed approach derives a direct data-driven discrete-time BBR algorithm which obtains a nonlinear stabilizing optimal control law that solves Problem \ref{problem:BBR}.
First, we construct an equivalent nonlinear optimization of Problem \ref{problem:BBR} via several transformations on the state and input decision variables. Next, we extract, via Pontryagin's Minimum Principle (PMP), a pair of first-order necessary conditions which allow for the extraction of the nonlinear control law from consideration of the costate dynamics. Point-to-point in time, these first-order conditions becoms linear matrix equalities which can be easily solved in an iterative manner. Considerations of the marginal sample autocorrelation allow for compact data-driven representations of these equalities. 

\paragraph{Reparametrization of P$_\text{BBR}$.}
Consider the parametrizations: ${G} \coloneqq \begin{bmatrix}   A & B & D  \end{bmatrix}$,  $I(K(x_t),x_t) \coloneqq \begin{bmatrix} \mathbb I_n & K^T(x_t) &x^T_t \otimes K^T(x_t)\end{bmatrix}^T $, $\mathbf G(K(x_t),x_t) \coloneqq I(K(x_t),x_t)  G$, and $N \coloneqq n +m +nm$ such that the augmented data vector $\xi_t \coloneqq \begin{bmatrix} x^T_{t} & u^T_{t} & (x_{t} \otimes u_{t})^T \end{bmatrix}^T \in \mathbb R^N$ with the propagation dynamics:
\begin{align}\label{eqn:bti_ltv}
\begin{split}
        \xi_{t} & =  \mathbf G(K(x_t),x_t) \xi_{t-1}, \quad t \in \mathbb N_+.
\end{split}
\end{align}
We use the propagation dynamics \eqref{eqn:bti_ltv} and the parameter $S_t \coloneqq \xi_t \xi^T_t$ to rewrite P$_\text{BBR}$ as the nonlinear optimization problem P$_1$.
 \begin{align}\label{eqn:problemP1}
    \text{P}_1: 
\begin{cases}
  &   \min\limits_{K(x_{t})}  {\hat{J}}_0^\infty \coloneqq   \sum\limits_{t=0}^{\infty} \mathbf{Tr}( \Lambda S_t)  \\
  &  \text{subject to:} \,    S_t  = \mathbf G(K(x_t),x_t)S_{t-1} \mathbf G^T(K(x_t),x_t), \\
  & \qquad \qquad  \; \; \rho(\mathbf G(K(x_t),x_t)) <1, \forall x_t, \\
  & \qquad \qquad  \; \; \lim\limits_{t \to \infty} S_t = 0,   t \in \mathbb N_+,
\end{cases}
\end{align}
where $\Lambda \coloneqq \texttt{blkdiag}(Q,R,F)$. Lemma \ref{lemma:equivalentproblems} proves that for a control law of the form $u_t = K(x_t) x_t$,  solving P$_1$ is equivalent to solving P$_\text{BBR}$.
\begin{lemma}\label{lemma:equivalentproblems}
Let $(K^\star_{\text{P}_\text{BBR}}(x_t),J_t^{t \star})$ and $(S_t^\star, K^\star_{\text{P}_1}(x_t),\hat J_t^{t \star})$ denote the respective optimal solution tuples to P$_\text{BBR}$ and P$_1$ at time step $t$. Given $u_t = K(x_t) x_t$, P$_1$ is equivalent to P$_\text{BBR}$ in the sense that $(K^\star_{\text{P}_\text{BBR}}(x_t),J_t^{t \star})=(K^\star_{\text{P}_1}(x_t),\hat{J}_t^{t \star})$ for all $x_t$, $t \in \mathbb N_+$.
\end{lemma}
\begin{proof}
The tuple $(x_{t-1}, u_{t-1},x_t)$ is known from the last state transition propagated by \eqref{eqn:bti_plant}. Hence, at $x_t$ under the  parametrizations $S_t$ and $\mathbf G(K(x_t),x_t)$,
\begin{equation}
    J_t^t = \begin{bmatrix}x^T_t  & u^T_t & (x_t \otimes u_t)^T \end{bmatrix} \Lambda \begin{bmatrix}x^T_t & u^T_t & (x_t \otimes u_t)^T \end{bmatrix}^T  = \mathbf{Tr}(\Lambda S_t) = \hat J_t^t,
\end{equation} and the dynamics in \eqref{eqn:bti_ltv} implies that
\begin{align}\label{eqn:plantconstraint}
  \begin{split}
        &   S_t  = \mathbf G(K(x_t),x_t)S_{t-1} \mathbf G^T(K(x_t),x_t),
  \end{split}
\end{align}
which is the equality constraint in~\eqref{eqn:problemP1}. The constraint $\lim_{t \to \infty} x_t = 0$ in P$_\text{BBR}$ directly implies $\lim_{t \to \infty} S_t = 0$ given $u_t = K(x_t) x_t$. At this point, we have proven that P$_\text{BBR}$ implies P$_1$ so that $\hat{J}_t^{t \star} \geq J_t^{t \star}$. Now, consider the constraint $\rho(\mathbf G(K,x_t))<1$ for all $t \in \mathbb N_+$. 
If $\rho(\mathbf G(K,x_t)) <1$ for all $x_t$ and $t \in \mathbb N_+$, then $S_{t-1}$ undergoes a contraction at each time step so that the asymptotic boundary condition on $S_t$ is satisfied and, immediately, $  \mathbf G(K(x_t),x_t) S_{t-1} \mathbf G^T(K(x_t),x_t) -S_{t-1} \preceq 0$. The inequality  $\mathbf G(K(x_t),x_t) S_{t-1} \mathbf G^T(K(x_t),x_t) - S_{t-1} \preceq 0$ has a unique limiting solution iff $S_{t-1} \succeq 0$ and $\rho(\mathbf G(K(x_t),x_t)) <1$, according to~\cite[Thm. 3.18, pg. 85]{Gu2012}. Accordingly, P$_1$ admits a unique solution $(S_t^\star,K^\star_{\text{P}_1}(x_t))$ at each time step $t$ where $K(x_t) = K^\star_{\text{P}_1}(x_t)$ is a feasible point of P$_\text{BBR}$ whose instantaneous cost is exactly ${J}_t^{t }(K^\star_{\text{P}_1}(x_t)) = \hat J_t^{t }(K^\star_{\text{P}_1}(x_t))$ at each $x_t$ and $t$. We conclude that  $K^\star_{\text{P}_1}(x_t)=K^\star_{\text{P}_\text{BBR}}(x_t))$ with $\hat{J}_t^{t \star}(K^\star_{\text{P}_1}(x_t)) =  J_t^{t \star}(K^\star_{\text{P}_1}(x_t))$ for all $x_t$, $t \in \mathbb N_+$. 
\end{proof}
Note that the constraint $\rho(\mathbf G(K(x_t),x_t))<1$ is difficult to handle since it is nonlinear, nonaffine, and not explicit in terms of $K(x_t)$. We show by Proposition~\ref{prop:replace-spectral} that we can safely remove the constraint $\rho(\mathbf G(K,x_t))<1$ since allowing $S_t \succeq 0$ is sufficient for solving P$_1$ without changing the optimal solution or cost function. We define and solve P$_2$.
\begin{proposition}\label{prop:replace-spectral}
The constraint $\rho(\mathbf G(K(x_t),x_t))<1$ can be replaced
with $S_t \succeq 0$ without changing the optimal solution or cost function of P$_1$.
\end{proposition}
\begin{proof}
The constraint $\rho(\mathbf G(K(x_t),x_t))<1$ means that $S_t$ undergoes a contraction at each time step so that the equality constraint in~\eqref{eqn:problemP1} is equivalent to
\begin{equation}\label{eqn:P1-equiv}
    \mathbf G(K(x_t),x_t) S_t \mathbf G^T(K(x_t),x_t) - S_t \preceq Z
\end{equation}
where $Z \succeq 0$ is some symmetric positive semi-definite matrix which can be written as $Z =M M^T$ given that there exists a $M \in \mathbb R^{N \times N}$. Then, the pair $(\mathbf G(K(x_t),x_t), \mathbf G(K(x_t),x_t) M)$ is stabilizable for any $K(x_t) \in \mathbb R^{m \times n}$ because the state-feedback gain $F = -  M^{\dagger}$ satisfies $\mathbf G(K(x_t),x_t) + \mathbf G(K(x_t),x_t) M F = \mathbf 0$. By~\cite[Thm. 3.18, pg. 85]{Gu2012}, if $(\mathbf G(K,x_t), \mathbf G(K(x_t),x_t) M)$ stabilizable, then~\eqref{eqn:P1-equiv} has a solution $S_t\succeq 0$ if and only if $\rho(\mathbf G(K(x_t),x_t)) < 1$. This shows the constraint $\rho(\mathbf G(K(x_t),x_t))<1$ can be safely replaced by $S_t\succeq 0$, which is satisfied by construction.
\end{proof}
\begin{align}\label{eqn:problemP2}
    \text{P}_2: 
\begin{cases}
  &   \min\limits_{K(x_{t})}  {\hat{J}}_0^\infty \coloneqq   \sum\limits_{t=0}^{\infty} \mathbf{Tr}( \Lambda S_t)  \\
  &  \text{subject to:} \,    S_t  = \mathbf G(K(x_t),x_t)S_{t-1} \mathbf G^T(K(x_t),x_t), \\
  & \qquad \qquad  \; \; S_t \succeq 0, \; \forall t, \\
  & \qquad \qquad  \; \; \lim\limits_{t \to \infty} S_t = 0,   t \in \mathbb N_+,
\end{cases}
\end{align}

P$_2$ is highly nonlinear and difficult to solve over any time horizon of any length $\geq 2$ (infinite or otherwise). Instead, we propose to solve P$_2$ in instantaneous time increments (point-to-point in time) for the state-dependent optimal control law $u_t = K(x_t) x_t$. Consider that at a fixed time step $t_k-1$, the tuple $(x_{t_k-1}, u_{t_k-1},x_{t_k})$ is known from the last state transition propagated by the bilinear system~\eqref{eqn:bti_plant}, i.e., $u_{t_k-1}$ was input to the bilinear system~\eqref{eqn:bti_plant} to drive $x_{t_k-1}$ to the next observed state $x_{t_k}$. Now, $S_{t_k-1}$ is exactly known and $\mathbf G(K(x_{t_k}),x_{t_k})$ is dependent only on the unknown decision variable $K(x_{t_k})$. Thus, the nonlinear nonconvex equality in~\eqref{eqn:problemP2} becomes a quadratic matrix equality, convex in $S_{t_k}$ and $K(x_{t_k})$. This analysis motivates us to consider the use of the Hamiltonian and PMP to find the best possible control input (at each point in time) that drives the bilinear system from one state to the next. 
\paragraph{The Hamiltonian and Pontryagin's Minimum Principle (PMP).}
Consider the Hamiltonian of the dynamic system in~\eqref{eqn:problemP2}:
\begin{align}\label{eqn:hamiltonian}
    \begin{split}
         \mathcal H_t \coloneqq \mathbf{Tr}(\Lambda S_t + P_t^T \mathbf G(K(x_t),x_t)S_{t-1} \mathbf G^T(K(x_t),x_t)),
    \end{split}
\end{align}
where $P_t \in \mathbb R^{N \times N}$ is the costate matrix at timestep $t$. From PMP, the first-order necessary conditions on $\mathcal H_t$ are\footnote{Eq. is shorthand for equation.} 
\begin{subequations}\label{eqn:firstorderconditions}
\begin{align}
 & \text{State Eq. : } \frac{\partial \mathcal H_t}{\partial P_t}  = \mathbf G(K(x_t),x_t)S_{t-1} \mathbf G^T(K(x_t),x_t))= S_t, \label{eqn:FOC_dynamics}  \\
& \text{Costate Eq. : }\frac{\partial \mathcal H_t}{\partial S_t}  = \Lambda + \mathbf G^T(K,x_t) P_t \mathbf G(K(x_t),x_t) = P_{t-1}, \label{eqn:FOC_costate}  \\
    & \text{Input Eq. : } \frac{\partial \mathcal H_t}{\partial K(x_t)} = 2  G S_{t-1} G^T   [P_t^{n \times (N-n)} L(x_t)    + K^T(x_t) U(P_{t},x_t)] =0, \label{eqn:FOC_input}\\
& \text{Boundary Condition: } \lim\limits_{t \to \infty} S_t =0, \label{eqn:FOC_boundary}  \\
 & \text{Transversality Constraint: }     \lim_{t\to \infty} S_t P_t \succeq 0,  \label{eqn:FOC_transversality} 
\end{align}
\end{subequations}
where $U(P_{t},x_t) \coloneqq L^T(x_t) P_t^{(N-n)  \times (N-n)} L(x_t)$, and $L(x_t) \coloneqq \begin{bmatrix} \mathbb I_m & x^T_t \otimes \mathbb I_m \end{bmatrix}^T$. In~\eqref{eqn:firstorderconditions}, PMP gives a pair of (backwards) iterative control law evaluation (costate update~\eqref{eqn:FOC_costate}) and improvement (control gain update~\eqref{eqn:FOC_input}) conditions. From~\eqref{eqn:FOC_input}, we can directly extract the control gain $K(x_t)$ in terms of the costate matrix $P_t$.
\begin{equation}\label{eqn:FOC_controlgain}
     K(x_t)  = - U(P_{t},x_t)^{\dagger} L^T(x_t) (P_t^{n \times (N-n)})^T.
\end{equation}
Solving~\eqref{eqn:firstorderconditions} and~\eqref{eqn:FOC_controlgain} is intractable due to the following difficulties. First, for $\mathbf G(K(x_t),x_t)$ known and independent of $x_{t}$ (e.g. time invariant or time-varying explicit only in $t$),~\eqref{eqn:FOC_costate} and~\eqref{eqn:FOC_input} can be easily solved (offline) by sweeping backwards in time \cite[Chapter 4]{bryson2017applied}. However, in our case, the costate matrix, $P_t$, which evolves backward in time, is cyclically dependent on $x_t$ which evolves forward in time under $u_t$ dependent on $P_t$.
To resolve the cyclic dependency, we propose to solve the following iterative numerical procedure, inspired by the procedure that solves the discrete-time State-Dependent Riccati Equation (DSDRE) in \citep{yoshida2012optimality}, in which the pointwise stabilizing optimal control gain is obtained by regarding all state-dependent matrices as \emph{constant} at each point in time $t$. At each timestep $t$, consider that we treat the state $x_t$ as a fixed constant parameter $\theta$ so that $\mathbf{G}(K(\theta),\theta)=\mathbf G(K(x_t),x_t)$. We solve the first-order necessary conditions~\eqref{eqn:FOC_costate},~\eqref{eqn:FOC_transversality}, and~\eqref{eqn:FOC_controlgain} that sweep a fictitious costate matrix $P_{t,k}$ backwards over some interval $k \in [0,k_f]$ under $\mathbf{G}(K(\theta),\theta)$ with $\theta$ fixed. We give these modified equations as
\begin{subequations}\label{eqn:firstorderconditions_iterate}
\begin{align}
&  \Lambda + \mathbf G^T(K_k,\theta) P_{t,k} \mathbf G(K_k(\theta),\theta) = P_{t,k-1},\label{eqn:FOC_costate_iterate}  \\
    &    K_k(\theta)  = - U(P_{t,k},\theta)^\dagger L^T(\theta) (P_{t,k}^{n \times (N-n)})^T, \label{eqn:FOC_input_iterate}     \; \\
  &   \lim_{k\to 0} S_t P_{t,k} \succeq 0, \; k \in [0,k_f],  \label{eqn:FOC_transversality_iterate} 
\end{align}
\end{subequations}
where, with some abuse of notation, $P_{t,k}$, $K_{k}(\theta)$, and $\mathbf G(K_k,\theta)$ are the respective $k$-th update of $P_t$, $K(\theta)$, and $\mathbf G(K(\theta),\theta)$. Implicitly, we recover the boundary constraint $P_{t,k_f}=\mathbf 0$ from~\eqref{eqn:FOC_transversality_iterate}. According to \citep{anderson2007}, $\lim_{k \to 0} P_{t,k}=P_t$ exists for all $t$ for stabilizability of $(\mathbf G^T(K_k,\theta),\Lambda^{1/2})$ so that $\lim_{k \to 0} K_{k}(\theta)=K(\theta) =K(x_t)$. 
\begin{proposition}\label{prop:controllability}
The pair $(\mathbf G^T(K_k(\theta),\theta),\Lambda^{1/2})$ is stabilizable for any $K_k(\theta)$ and $\theta$.
\end{proposition}
\begin{proof}
The pair $(\mathbf G^T(K_k(\theta),\theta),\Lambda^\frac{1}{2})$ is stabilizable if and only if $(\mathbf G^T_\Omega ,\Omega^{-T}\Lambda^\frac{1}{2})$ is stabilizable where $\mathbf G_\Omega \triangleq\Omega^{-1}\mathbf G(K_k(\theta))\Omega = \begin{bmatrix}
      G I(K_k(\theta),\theta) & B & D \\ \mathbf 0 & \mathbf 0 & \mathbf 0
   \end{bmatrix}$, and the nonsingular matrix $\Omega \triangleq $ $  \begin{array}{@{}c c@{}}
   \begin{bmatrix}
      I(K_k(\theta),\theta) \,\vrule
      & \begin{array}{c@{}} \mathbf 0  \\  \mathbb I_{m+nm} \end{array} 
   \end{bmatrix}  \end{array}  $ is a similarity transform with its inverse $\Omega^{-1} = \begin{array}{@{}c c@{}}
   \begin{bmatrix}
      I(-K_k(\theta),\theta) \,\vrule
      & \begin{array}{c@{}} \mathbf 0  \\  \mathbb I_{m+nm} \end{array} 
   \end{bmatrix}  \end{array}  $ and $\Omega^{-1}\Lambda^\frac{1}{2} = \Lambda^\frac{1}{2}$. Note that the stabilizability of $(\mathbf G^T_\Omega,\Lambda^\frac{1}{2})$ is equivalent to the detectability of $(\Lambda^\frac{1}{2},\mathbf G_\Omega)$. Under Assumption~\ref{assumption:detectability}, $(\Lambda^\frac{1}{2},\mathbf G_\Omega)$ is detectable.
\end{proof}
\begin{remark}
\eqref{eqn:firstorderconditions_iterate} is meant to be solved iteratively backwards in $k$. Given the boundary constraint $P_{t,k_f}= \mathbf 0$, we first solve~\eqref{eqn:FOC_input_iterate} for $K_{k_f}(\theta)$ which is then used in~\eqref{eqn:FOC_costate_iterate}. It is easy to see that at each $k$, the costate dynamics~\eqref{eqn:FOC_costate_iterate} is a LME, solvable for $P_{t,k}$.
\end{remark}
\begin{remark}
By assuming $x_t=\theta$ is fixed, the equations in~\eqref{eqn:firstorderconditions_iterate} with a boundary condition $P_{t,k_f}=\mathbf 0$ are effectively solving the two point boundary value problem (TPBVP) that iterates over the control policy space to find a linear control gain for a system characterized by $\mathbf{G}(K(\theta),\theta)$. However, solving~\eqref{eqn:FOC_costate_iterate} with $P_{t,k_f}$ does not satisfy~\eqref{eqn:FOC_costate} for optimality. At best, we can only recover a control gain solution that is locally optimal at the state $x_t$.
\end{remark}
The second difficulty is that the system model $G$ is unknown. If the system model $G$ was known, we would be able to solve~\eqref{eqn:firstorderconditions_iterate}. Note that~\eqref{eqn:FOC_input_iterate} is already in direct form. 

\RestyleAlgo{ruled}
\begin{algorithm}[!t]
 Given $\Lambda$, $k_f$ large, a tolerance $\epsilon >0$, $r$, and $t=0$ \\
 \Do{\texttt{True}}{
     Compute $\mathcal R_{\xi}^0$ as in \eqref{eqn:autocorrelation}\\
        \For{\texttt{each trajectory $i  \in \{1,\ldots,r\}$}}{
    Set $P_{t,k_f}= \mathbf 0$ and $\theta=x_t$ \\
    \Do{$\lVert P_{t,k} - P_{t,k-1}\lVert _2\leq \epsilon$}{
    Update $K_k(\theta)$ using \eqref{eqn:FOC_input_iterate} \\
    Solve (\ref{eqn:FOC_costate_iterate_modelfree}) for ${P}_{t,k-1}$   \\
    Set $k \leftarrow k-1$
  } 
  $u_t \gets K(x_t) x_t$ \\
  }   $t\gets t+1$
    } 
    \caption{Direct Data-Driven BBR Algorithm}
 \label{alg:onlineLQ}
 \end{algorithm}
\paragraph{Direct Data-Driven Representation.} Without knowledge on $G$, how do we solve~\eqref{eqn:firstorderconditions_iterate}  in a data-driven manner? In this section, we transform~\eqref{eqn:firstorderconditions_iterate} into a data-driven form. We assume that we have access to the bilinear system~\eqref{eqn:bti_plant} so that we can collect $r(>0)$ trajectories. Define $\xi_{t,i}$ be the augmented data vector $\xi_t$ observed at time step $t$ from the $i$-th trajectory, where $i \in [1,\ldots,r]$. Now, define the lag-$0$ marginal sample autocorrelation of the data vector $\xi_t$ as
\begin{equation}\label{eqn:autocorrelation}
   \mathcal  R^0_\xi \coloneqq \frac{1}{r} \sum_{i=1}^r \left(\frac{1}{j}\sum_{t}^{t+j-1} \xi_{t,i} \xi^T_{t,i}\right) ,
\end{equation}
computed over the $r$ trajectories of length $j \in \mathbb N_+$. For $\mathcal R^0_\xi \succ 0$, we can use $\mathcal R^0_\xi$ as a similarity transform to obtain the equivalent data-driven form of~\eqref{eqn:FOC_costate_iterate} as:
\begin{align}\label{eqn:FOC_costate_iterate_modelfree}
    \begin{split}
      &          \mathcal R_\xi^0\left(\Lambda -P_{t,k-1}\right) \mathcal R_\xi^0 + \mathbf V(K_k(\theta),\theta) P_{t,k} \mathbf V^T(K_k(\theta),\theta) = \mathbf 0,
    \end{split}
\end{align}
where
$
    \mathbf V(K_k(\theta),\theta) \coloneqq  \frac{1}{r} \sum_{i=1}^r \left(\frac{1}{j}\sum_{t}^{t+j-1} \xi_{t,i} x^T_{t+1,i} I^T(K_k(\theta),\theta)\right)$.  A natural question that arises is: is it possible to obtain $\mathcal R_\xi^0 \succ 0$ for the bilinear system~\eqref{eqn:bti_plant} under Assumption~\ref{assumption:span}? We present two methods for satisfying this condition.
    \begin{enumerate}
        \item \textbf{Varying the initial conditions for multiple trajectories:} One way in which to satisfy this condition is to let the number of trajectories $r \geq N$ and vary the initial conditions $\xi_{0,i}$ for $i \in [1,\ldots,r]$.    
\item \textbf{Obtain Persistently-exciting data of order $N$:} Utilize~\citep[Algorithm 1]{yuan2022data} to obtain $N$-persisently excited data so that $\mathcal R_\xi^0 \succ 0$. Note that this method is suitable for the case in which we can only obtain a single system trajectory $r=1$.
    \end{enumerate}
    
 The proposed direct data-driven BBR algorithm is presented in Algorithm~\ref{alg:onlineLQ}.
 \begin{remark}
 At each time step $t$, Algorithm~\ref{alg:onlineLQ} observes a worst-case polynomial time complexity in $N$. 
\end{remark} 

\section{Numerical Simulations}\label{section:examples_BBR}
In this section, we demonstrate the efficacy of the proposed direct data-driven BBR Algorithm (Algorithm~\ref{alg:onlineLQ}) via two numerical examples. Note that $x_t(j)$ denotes the $j$-th component of the state vector $x_t$.

\textit{{Example~\ref{section:examples_BBR}.1. Fission in Nuclear Reactors~\citep{Mohler1970}:}}
With a neutron population level $x_t(1)$ and neutron precursor $x_t(2)$, the discretized point neutron dynamics from nuclear fission is described as a  bilinear model as in (\ref{eqn:bti_plant}) with 
\begin{align}\label{eqn:nuclearfission}
      \begin{split}
          & A = \begin{bmatrix} \frac{1-T_s}{l} & \lambda T_s \\ 0 & 1-\lambda T_s  \end{bmatrix},   D_1 = \begin{bmatrix} \frac{1-\beta}{l} \\ \frac{\beta}{l}  \end{bmatrix} T_s,B = D_2 = \mathbf 0,
      \end{split}
  \end{align}
  where $D = [D_1,D_2]$ and $T_s=0.001$s is the sampling time. The precursor portion, $\beta=0.157 \times 10^{-3}$, has a neutron generation time $l=8.36 \times 10^{-4}$s and a decay constant $\lambda=0.0120 \text{s}^{-1}$. The input $u_t \in \mathbb R$ is the adjustable neutron multiplicative factor used to control the rate of the nuclear chain reaction. We attempt to quickly stabilize the neutron population by choosing $Q = \text{diag}(1,0)$, $R=0.1$, $F=\mathbb I_2$. Note that the system~\eqref{eqn:nuclearfission} satisfies Assumption \ref{assumption:span}. 

 We choose to vary the initial conditions by selecting the set of initial conditions: $x_{0,i} \sim \mathcal N(0,10 \mathbb I_n)$ 
 and $u_{0,i} \sim \mathcal N(0,\mathbb I_m)$ for $i = \{1,\ldots,r\}$ such that $\mathcal R^0_\xi \succ 0$.

\textit{Example~\ref{section:examples_BBR}.2. Hydraulic Rotary Multi-motor System \citep{shaker2013interaction}:}
The discretized  (with sampling time $T_s = 0.001$) hydraulic rotary system with two motors can be written as in~\eqref{eqn:bti_plant} with
\begin{align}\label{eqn:example2}
      \begin{split}
         & A = \mathrm{diag} (0.99997,-0.99997,0.99997),  B = \mathrm{diag} (200,6,10),  \\ & D_2 =\begin{bmatrix} 0 & -0.00007 & 0\\
0 & 0 & 0\\
0 & 0 & 0\end{bmatrix},    D_1 = \begin{bmatrix} 0 & 0 & 0 \\
0 & 0 & 0 \\
0 & 0.03 & 0.03\end{bmatrix}, \\ & D_3 =\begin{bmatrix} 0 & 0 & -0.00007 \\
0 & 0 & 0 \\
0 & 0 & 0\end{bmatrix}, D = \begin{bmatrix}
    D_1, D_2, D_3
\end{bmatrix}.
      \end{split}
  \end{align}
The state vector is comprised of the pressure level in the line-network, $x_t(1)$, and the speed of the first and second motor, $x_t(2)$ and $x_t(3)$, respectively. The inputs are the displacement volumes of the hydraulic pump, the first motor, and second motor, respectively. We set $Q= \mathbb I_3$, $R= \mathbb I_3$, $F=\mathbb I_{6}$. Note that the system~\eqref{eqn:example2} satisfies Assumption \ref{assumption:span}.

 We choose to vary the initial conditions by selecting the set of initial conditions: $x_{0,i} \sim \mathcal N(0,10 \mathbb I_n)$ 
 and $u_{0,i} \sim \mathcal N(0,\mathbb I_m)$ for $i = \{1,\ldots,r\}$ such that $\mathcal R^0_\xi \succ 0$.
 \begin{figure}[t!]
\centering
\includegraphics[width=0.915\textwidth]{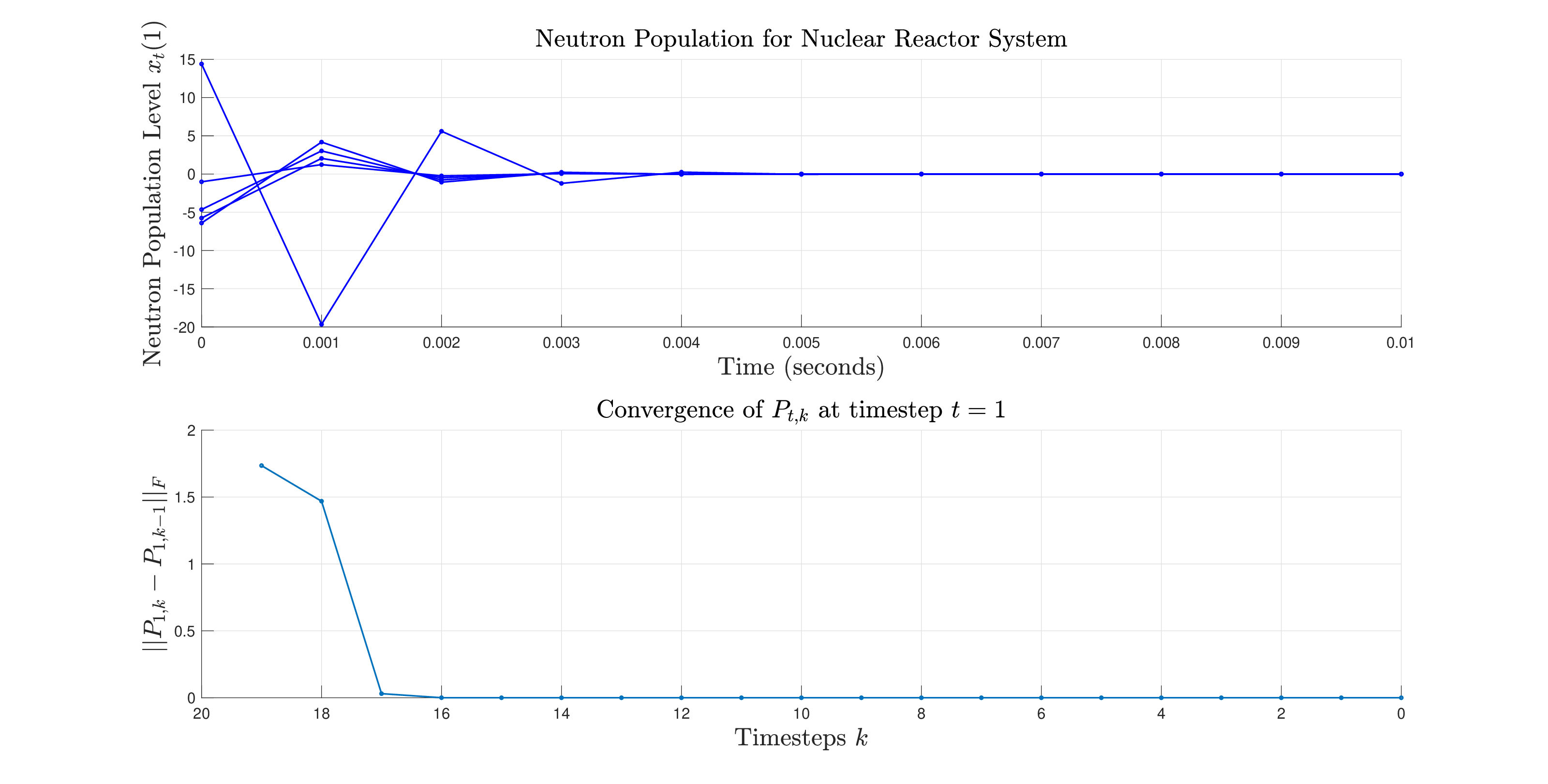}
  \caption{Performance of Algorithm~\ref{alg:onlineLQ} for a Nuclear Fission Reactor system~\cite{Mohler1970}.}
                  \label{fig:examples_a}
\end{figure}
\begin{figure}
\centering
\includegraphics[width=0.9\textwidth]{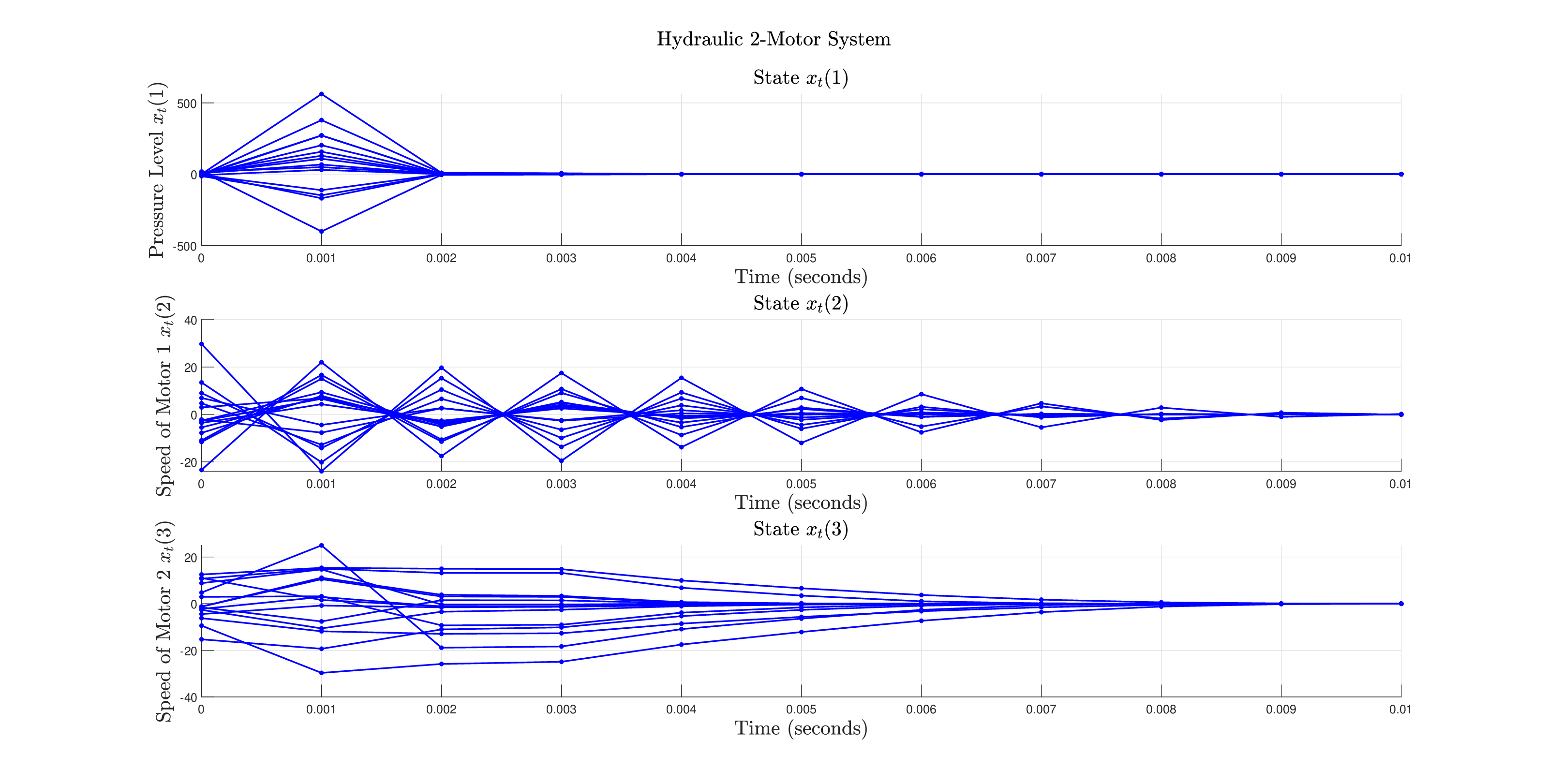}
        \caption{ Performance of Algorithm~\ref{alg:onlineLQ} for a 2-motor Hydraulic system~\cite{shaker2013interaction}.}
        \label{fig:examples_b}
\end{figure}
\paragraph{Discussion of Results:}
Figures~\ref{fig:examples_a} and~\ref{fig:examples_b} show the performance results of Algorithm~\ref{alg:onlineLQ} for the Nuclear Fission Reactor system and the Multi-motor Hydraulic system, respectively. The upper subplot of Fig.~\ref{fig:examples_a} and the plot of Fig.~\ref{fig:examples_b} demonstrate that, for varying initial conditions of $r=N$ trajectories ($N$ is appropriate to the respective system), Algorithm~\ref{alg:onlineLQ} is able to steer the designated state of the system to the origin. Furthermore, the lower subplot in Fig.~\ref{fig:examples_a} shows a snapshot of the convergence of the costate matrix $P_{t,k}$ at $k \to 0$ at timestep $t=1$ for a single trajectory of a Nuclear Fission Reactor system~\cite{Mohler1970}. We see that $\lim_{k \to 0} P_{1,k}=P_1$ for some value $P_1$. 

\section{Conclusions}\label{section:conclusion}
In this paper, we developed a data-driven optimal control algorithm for a new class of optimization problem called the Bilinear Biquadratic Regulator that learns a state-dependent optimal control law for an unknown bilinear system. To improve the computational tractability of the BBR, we derived an equivalent nonlinear optimization problem from which a system of data-encoded linear matrix equalities were solved in a point-to-point manner to obtain the optimal control law.  Through two numerical examples, we demonstrated the efficacy of the proposed algorithm. Future work will study the extension of the proposed algorithm to the noisy unknown bilinear system.

\bibliography{l4dc2024-sample}

\end{document}